\documentclass[reprint,aps,twocolumn,superscriptaddress,showpacs,notitlepage,10pt]{revtex4-2}

\usepackage{amsmath,amssymb,amsfonts,amstext,amsthm}
\usepackage{graphicx,breakurl,color,setspace}
\usepackage{mathtools}
\usepackage{dsfont}
\usepackage[T1]{fontenc}
\usepackage[latin9]{inputenc}
\usepackage[normalem]{ulem}
\usepackage[unicode=true,pdfusetitle,pdfborder={0 0 1}]{hyperref}
\usepackage{xcolor}

\hypersetup{
    colorlinks,
    linkcolor=[RGB]{243 68 110},
    citecolor=[RGB]{0 160 63},
    urlcolor=[RGB]{57 115 239}
}

\usepackage{upgreek}
\usepackage{bm}
\usepackage{mathrsfs}

\begin{document}

\title{Transversal Logical Clifford gates on rotated surface codes with reconfigurable neutral atom arrays} 

\author{Zihan Chen}


\affiliation{
Hefei National Research Center for Physical Sciences at the Microscale and School of Physical Sciences,
University of Science and Technology of China, Hefei 230026, China}
\affiliation{
Shanghai Research Center for Quantum Science and CAS Center for Excellence in Quantum Information and Quantum Physics,
University of Science and Technology of China, Shanghai 201315, China}
\affiliation{
Hefei National Laboratory, University of Science and Technology of China, Hefei 230088, China}

\author{Mingcheng Chen}

\affiliation{
Hefei National Research Center for Physical Sciences at the Microscale and School of Physical Sciences,
University of Science and Technology of China, Hefei 230026, China}
\affiliation{
Shanghai Research Center for Quantum Science and CAS Center for Excellence in Quantum Information and Quantum Physics,
University of Science and Technology of China, Shanghai 201315, China}
\affiliation{
Hefei National Laboratory, University of Science and Technology of China, Hefei 230088, China}

\author{Chaoyang Lu}

\affiliation{
Hefei National Research Center for Physical Sciences at the Microscale and School of Physical Sciences,
University of Science and Technology of China, Hefei 230026, China}
\affiliation{
Shanghai Research Center for Quantum Science and CAS Center for Excellence in Quantum Information and Quantum Physics,
University of Science and Technology of China, Shanghai 201315, China}
\affiliation{
Hefei National Laboratory, University of Science and Technology of China, Hefei 230088, China}

\author{Jianwei Pan}

\affiliation{
Hefei National Research Center for Physical Sciences at the Microscale and School of Physical Sciences,
University of Science and Technology of China, Hefei 230026, China}
\affiliation{
Shanghai Research Center for Quantum Science and CAS Center for Excellence in Quantum Information and Quantum Physics,
University of Science and Technology of China, Shanghai 201315, China}
\affiliation{
Hefei National Laboratory, University of Science and Technology of China, Hefei 230088, China}

\date{\today}

\begin{abstract}
We propose hardware-efficient schemes for implementing logical H and S gates transversally on rotated surface codes with reconfigurable neutral atom arrays. For logical H gates, we develop a simple strategy to rotate code patches efficiently with two sets of 2D-acousto-optic deflectors (2D-AODs). Our protocol for logical S gates utilizes the time-dynamics of the data and ancilla qubits during syndrome extraction (SE). In particular, we break away from traditional schemes where transversal logical gates take place between two SE rounds and instead embed our fold-transversal logical operation \textit{inside} a single SE round, leveraging the fact that data and ancilla qubits can be morphed to an \textit{unrotated} surface code state at \textit{half-cycle}. Under circuit noise, we observe the performance of our S gate protocol is on par with the quantum memory. Together with transversal logical CNOT gates, our protocols complete a transversal logical Clifford gate set on rotated surface codes and admit efficient implementation on neutral atom array platforms.   
\end{abstract}

\maketitle

Quantum error correction (QEC) is necessary, given the state-of-the-art physical error rates, for achieving a low enough logical error rate to run quantum algorithms at a practically relevant scale~\cite{gidney_how_to_factor_2021,kivlichan_improved_2020,campbell_early_2022}. A popular QEC code familiy is the rotated surface codes~\cite{horsman_rotated_surgery_2012} which have high circuit noise threshold $\sim 1\%$ and are now one of the focuses of experimental efforts in QEC across platforms~\cite{krinner_realizing_2022,google_quantum_ai_suppressing_2023,bluvstein_logical_2023,acharya2024quantumerrorcorrectionsurface}. In particular, most recently, the rotated surface code memory is  demonstrated definitively below threshold for the first time with superconducting qubits and the logical qubit of a $d=7$ rotated surface code is observed to preserve information for more than two times longer than a physical qubit~\cite{acharya2024quantumerrorcorrectionsurface}.

To perform fault-tolerant quantum computation (FTQC) on quantum systems with only local connectivity, lattice surgery~\cite{horsman_rotated_surgery_2012,litinski_game_2019} is a viable strategy for logical operations but requires extra ancilla patches and $O(d)$ time overhead for code distance $d$.  With the help of (effective) long-range connectivity, logical CNOT gates on rotated surface codes are recently demonstrated~\cite{bluvstein_logical_2023} for $d=7$ with neutral atom arrays, a nascent platform featuring scalability~\cite{manetsch_tweezer_2024}, high-fidelity operations~\cite{bluvstein_logical_2023,Yb_spectroscopy_2024,Sr_benchmarking_2024} and parallel coherent atom transport~\cite{bluvstein_quantum_coherent_transport_2022}.

It is thus tempting to consider how to achieve more efficient logical operations on rotated surface codes by utilizing the qubit reconfigurability of neutral atom array platforms. For logical Clifford gates, with a generating set $\{\mathrm{S},\mathrm{H},\mathrm{CNOT}\}$, it is well known that logical $\mathrm{H}$ and $\mathrm{CNOT}$ gates can be implemented transversally on rotated surface codes given long-range connectivity. To the best of our knowledge, it was unknown (prior to this work) how to do transversal $\mathrm{S}$ gates on rotated surface codes despite a series of works that improve the efficiency of the $\mathrm{S}$ gate under the constraint of local connectivity~\cite{fowler_surface_2012,brown_poking_2017,chamberland_universal_2022,bombin_logical_2023,gidney_inplace_2024}. In contrast, \textit{unrotated} surface codes admit transversal implementation of all three logical Clifford gates in the generating set above, albeit requiring almost twice as many code qubits as the rotated surface codes for the same code distance~\cite{moussa_transversal_2016,breuckmann_fold_transversal_2024,zhou_algorithmic_2024}. Moreover, considering hardware limitations, implementing the transversal H gate can be cumbersome as it requires patch rotation which is not naturally supported by the reconfiguration device, 2D-acousto-optic deflectors (2D-AOD), currently used on neutral atom array platforms~\cite{bluvstein_quantum_coherent_transport_2022,xu_constant_overhead_2023}.

In this work, we complete the transversal logical Clifford gate set on rotated surface codes by proposing a transversal S gate protocol. Moreover, we make both the transversal H gate and our transversal S gate protocol efficiently implementable with neutral atom arrays. We propose to use an extra set of 2D-AOD, diagonally aligned, together a horizontally aligned set, to implement patch rotation for transversal H gates. For logical $\mathrm{S}$ gates, we break away from the conventional static point of view where transversal gates act on data qubits between two syndrome extraction (SE) rounds. Instead, we embed our transversal operation inside a single SE round, building on a previous observation that data qubits and ancilla qubits can be \textit{morphed} to an \textit{unrotated} surface code state during an SE round~\cite{mcewen_relaxing_2023}. We also numerically demonstrate the logical error rate of our $\mathrm{S}$ gate protocol is on the same level with the (rotated surface code) quantum memory under circuit noise.

\begin{figure*}
    \centering
    \includegraphics[width=\textwidth]{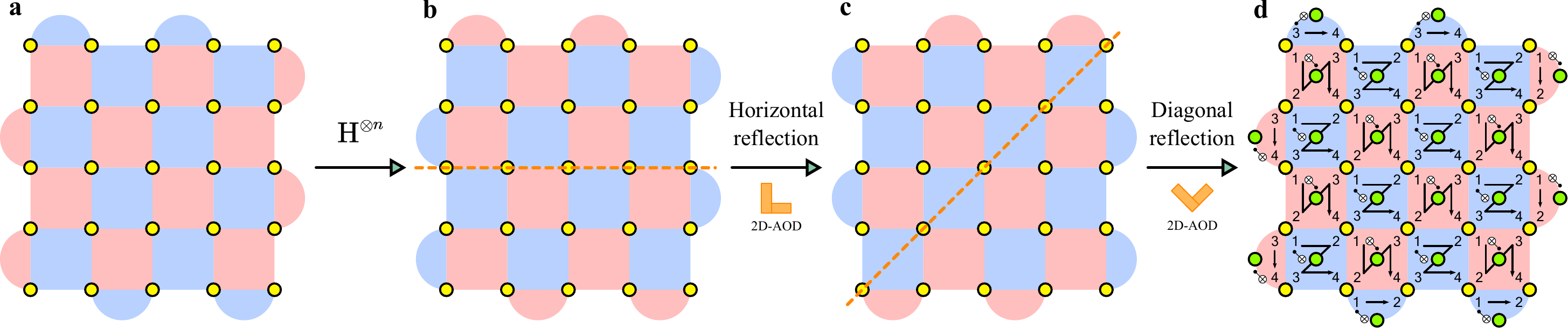}
    \caption{Transversal H gate protocol via patch rotation and the SE circuit. (\textbf{a}-\textbf{d}) Evolution of code stabilizers throughout the transversal H gate protocol. Light red (blue) tiles indicate X (Z) stabilizers. Data (ancilla) qubits are indicated by yellow (green) circles. We start with a rotated surface code in (\textbf{a}). By applying H gates over all data qubits, we perform a logical $\mathrm{H}$ gate but at the same time change all X stabilizers to Z stabilizers and vice versa as shown in (\textbf{b}). A horizontal reflection (via a horizontally aligned 2D-AOD) along the dashed line in (\textbf{b}) followed by a diagonal reflection (via a diagonally aligned 2D-AOD) along the dashed line in (\textbf{c}) would bring the stabilizer pattern back to what we start with, as shown in (\textbf{d}). In addition to the stabilizers, we also illustrate the SE circuit with 4 layers of CNOT gates in (\textbf{d}). Notice that X and Z ancilla qubits live on different diagonal lines. Thus, with a diagonally aligned 2D-AOD, we can address simultaneously all X (Z) ancilla qubits without affecting other qubits, which is helpful for implementing the CNOT gates during SE.}
    \label{fig: H gate}
\end{figure*}

\textit{Transversal H gate and patch rotation.} Transversal H gates on all data qubits followed by a $\pi/2$ patch rotation implement a logical H gate on a rotated surface code. For 2D neutral atom arrays, mid-circuit coherent atom rearrangement is implemented (in current experiments) via a 2D-AOD which can translationally move a rectangular grid of atoms but does not naturally support patch rotation~\cite{bluvstein_quantum_coherent_transport_2022,bluvstein_logical_2023}. In the following, we propose a hardware-efficient way to implement patch rotation using two set of 2D-AODs, one aligned horizontally and the other aligned diagonally. We notice that a rotation is equivalent to two successive reflections, each of which can be efficiently implemented via a properly-aligned 2D-AOD using the divide-and-conquer algorithm~\cite{xu_constant_overhead_2023}. Thus, using a 2D-AOD (horizontally-aligned) for horizontal reflection and another (diagonally-aligned) for diagonal reflection, we can efficiently implement the $\pi/2$ patch rotation (Fig.~\ref{fig: H gate}). Moreover, each set of 2D-AOD helps experimental implementation of FTQC on rotated surface codes in other important ways. The horizontally-aligned 2D-AOD is already used for selecting, addressing and moving rectangular code patches in experiment~\cite{bluvstein_logical_2023}. More surprisingly, a diagonally-aligned 2D-AOD actually helps with SE. As is shown in Fig.~\ref{fig: H gate}(\textbf{d}), the compact layout of the SE circuit requires different CNOT gate orderings for X and Z stabilizer measurements respectively. We call ancilla qubits for X (Z) stabilizer measurements as X (Z) ancilla qubits. Then to efficiently implement the SE circuit, we need to be able to address X and Z ancilla qubits separately. Given the qubit arrangement in Fig.~\ref{fig: H gate}(\textbf{d}), both X and Z ancilla qubits have a checkerboard pattern. Selecting all X ancilla qubits with a horizontally-aligned 2D-AOD will also unavoidably select all Z ancilla qubits (except for a few at the boundary), which is inconvenient for SE. On the other hand, with a diagonally-aligned 2D-AOD, we can address all X (Z) ancilla qubits without affecting any other qubits since X and Z ancilla qubits are on different diagonal lines. In addition to helping with SE, the diagonally-aligned 2D-AOD is also needed for our $\mathrm{S}$ gate protocol which we will elaborate on in the following.

\begin{figure}[!htbp]
    \centering
    \includegraphics[width=\columnwidth]{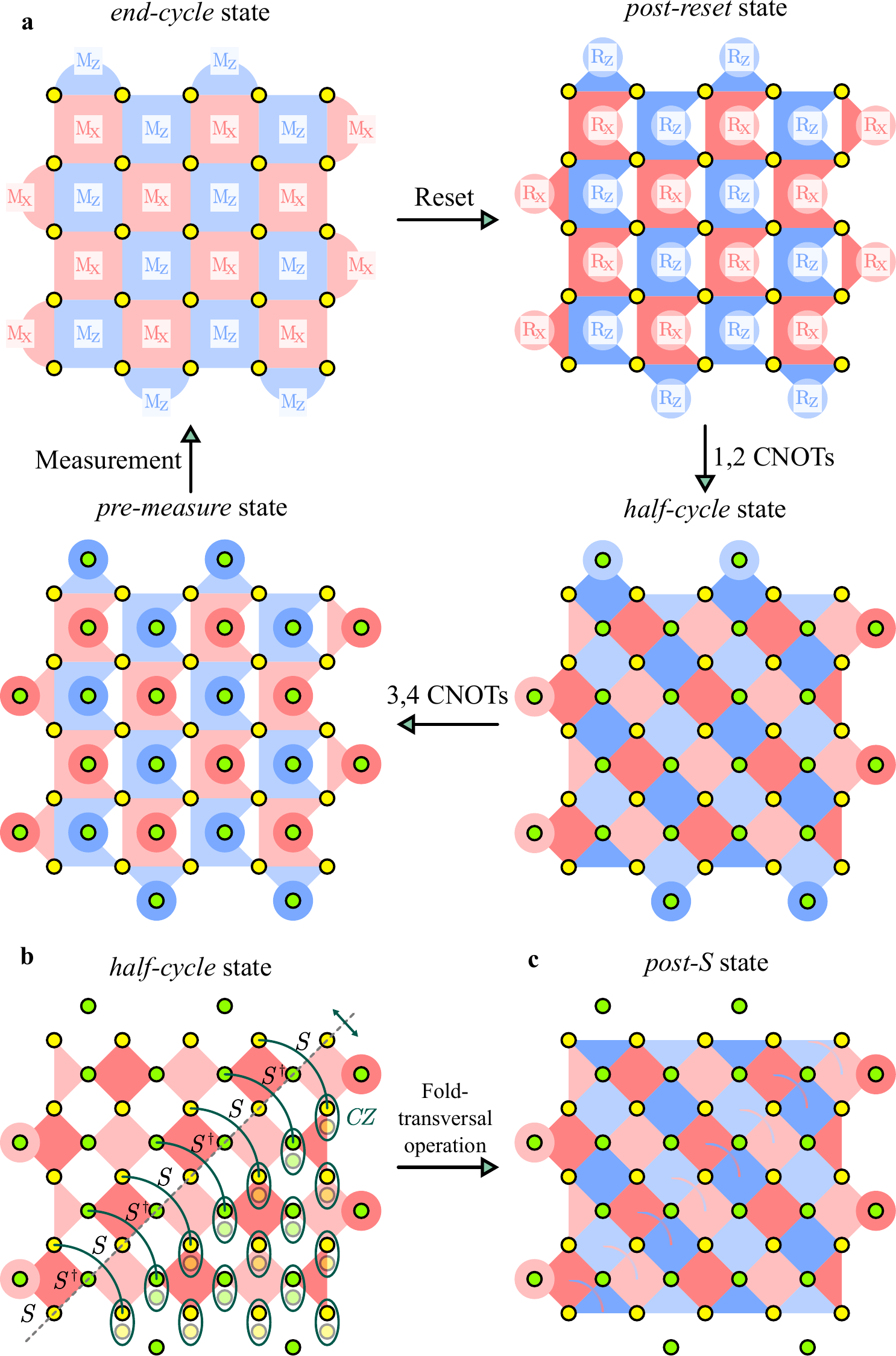}
    \caption{$\mathrm{S}$ gate protocol based on the \textit{morphing} effect of the SE circuit. (\textbf{a}) \textit{Mid-cycle} states during SE~\cite{mcewen_relaxing_2023}. Red and light red are used for Pauli X operators. Blue and light blue are used for Pauli Z operators. In particular, stabilizers with lighter tints are induced solely by reset locations in the current SE round.  (\textbf{b-c}) Fold-transversal operation and propagation of X stabilizers. No Z stabilizer is illustrated in (\textbf{b}) nor (\textbf{c}).  As is shown in (\textbf{b}), we apply a fold-transversal operation~\cite{moussa_transversal_2016,breuckmann_fold_transversal_2024} on the bulk of the \textit{half-cycle} state.  Due to the fold-transversal operation, each X stabilizer on the \textit{half-cycle} state in the bulk (shown in (\textbf{b})) propagates to the product of itself and its mirrored Z tile on the \textit{post-S} state (shown in (\textbf{c})). It is interesting to notice that the pattern in (\textbf{c}) is slightly different from the \textit{half-cycle} state in (\textbf{a}) at the top and bottom boundary. Such difference would introduce extra complications for detector construction (Fig.~\ref{fig: detectors}(\textbf{b})). }   
    \label{fig: mid cycle states and s gate protocol}
\end{figure}

\textit{Transversal S gate protocol.} We begin by briefly reviewing the (cyclic) \textit{morphing} of stabilizers on both the data and ancilla qubits during SE rounds, following~\cite{mcewen_relaxing_2023}. As in Fig.~\ref{fig: mid cycle states and s gate protocol}(\textbf{a}), at the end of the $i$-th SE round, after all ancilla qubits are measured, the data qubits (unmeasured) constitute the \textit{end-cycle} state (at round $i$) which is a rotated surface code state. At the beginning of round $i+1$, ancilla qubits are reset and the resulting state on all data and ancilla qubits is called the \textit{post-reset} state.  After two steps of CNOT gate operations, the transformed state, named the \textit{half-cycle} state, is simply an \emph{unrotated} surface code state along with a few unentangled qubits at the boundary. The next two steps of CNOT gates and ancilla measurements will \textit{morph} the \textit{half-cycle} state back to a rotated surface code state, which is the \textit{end-cycle} state at round $(i+1)$ (see Fig.~\ref{fig: mid cycle states and s gate protocol}(\textbf{a})). 

The crux of our S gate protocol is to apply the fold-transversal S gate~\cite{moussa_transversal_2016,breuckmann_fold_transversal_2024}  for \textit{unrotated} surface codes on the \textit{half-cycle} state. To be more specific, our S gate protocol is embedded in a single SE round $i_{\mathrm{S}}$ and consists of the following procedures:
\begin{itemize}
    \item [1.] At the beginning of round $i_{\mathrm{S}}$, reset the ancilla qubits and carry out the 1-2  CNOT layers according to the SE circuit in Fig.~\ref{fig: H gate}(\textbf{d}). 
    \item [2.] Perform the fold-transversal S gate on the \textit{half-cycle} state according to Fig.~\ref{fig: mid cycle states and s gate protocol}(\textbf{c}). More specifically, for qubits in the bulk, apply $\mathrm{S}$ and $\mathrm{S}^{\dagger}$ alternatingly on qubits along the diagonal line and $\mathrm{CZ}$ gates for each pair of qubits that are mirrored to each other along the diagonal line. 
    \item [3.] To return to a rotated surface code state, continue to perform the 3-4 CNOT layers and measure the ancilla qubits according to the SE circuit.    
\end{itemize}
It is straightforward to verify the above procedures realize a logical S gate on the rotated surface code and at the same time obtain a single round of SE results. We call each SE round that implements our $\mathrm{S}$ gate protocol as an $\mathrm{S}$-SE round and the conventional SE rounds as $\mathrm{I}$-SE rounds.

\textit{Detectors.} To signal the presence of errors in a circuit, we need to construct detectors which are checks for circuit errors. A detector of a circuit is a set of measurement results whose parity sum has a fixed value as long as the circuit is noiseless.  Constructing detectors for conventional transversal logical gates (that act on data qubits and occur between two SE rounds) can be done straight-forwardly by tracking the code stabilizers through the transversal operations~\cite{wan2024iterativetransversalcnotdecoder,sahay2024errorcorrectiontransversalcnot,zhou_algorithmic_2024}. On the other hand, as our mid-round fold-transversal operation acts on both data qubits and ancilla qubits, constructing detectors in our $\mathrm{S}$ gate protocol requires circuit-level perspectives~\cite{mcewen_relaxing_2023,gottesman_opportunities_2022,delfosse_spacetime_2023,beverland_fault_2024}. We rely on the detecting region formalism~\cite{mcewen_relaxing_2023} in this work and present how detectors can be constructed in the following. Notice that reset locations generate stabilizers (with fixed signs) which propagate deterministically through unitary gates. Thus, for a set of reset locations $\mathcal{R}$, if there is a set $\mathcal{M}$ of measurements that jointly measure the determinstically propagated stabilizer generated by $\mathcal{R}$, then $\mathcal{M}$ is a detector (induced by $\mathcal{R}$).  (See SM~\cite{supplimental} for a more detailed and precise description.) This perspective essentially echos the framework of detecting regions as the propagation of the stabilizer generated by $\mathcal{R}$ carves out the detecting region corresponding to detector $\mathcal{M}$.  We describe the construction of detectors for our $\mathrm{S}$ gate protocol in the following.

\begin{figure}[t]
    \centering
    \includegraphics[width=\columnwidth]{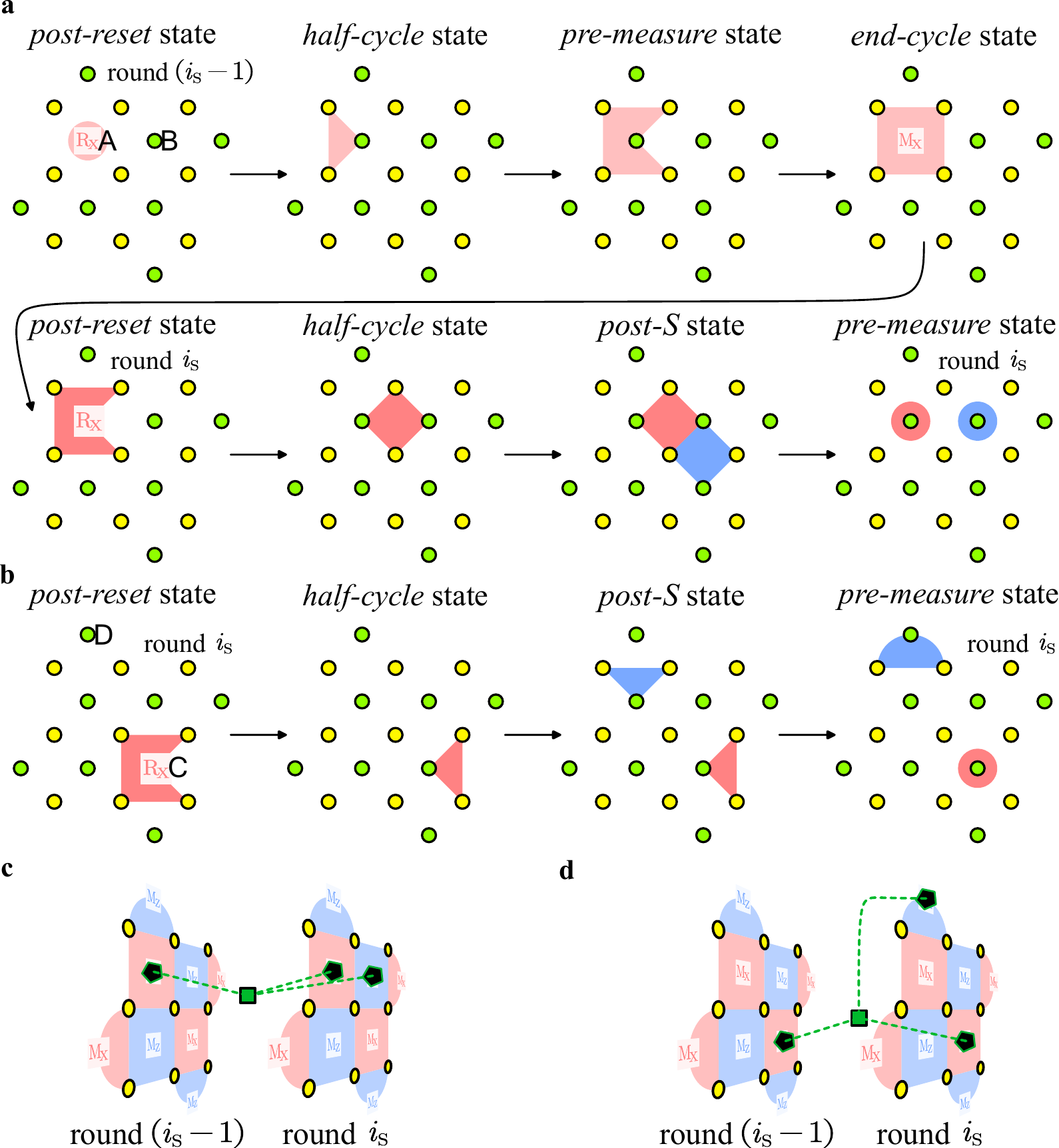}
    \caption{Construction of X detectors. Four ancilla qubits of interest are labeled by A-D respectively. (\textbf{a}) Propagation of the stabilizer created by two X reset locations $\mathsf{R}_{\mathrm{X}}(\mathrm{A},i_{\mathrm{S}}-1)$ and $\mathsf{R}_{\mathrm{X}}(\mathrm{A},i_{\mathrm{S}})$. The \textit{post-S} state is the state immediately after the fold-transversal operation. (\textbf{b}) Propagation of the stabilizer created by reset locations $\mathsf{R}_{\mathrm{X}}(\mathrm{C},i_{\mathrm{S}}-1)$ and $\mathsf{R}_{\mathrm{X}}(\mathrm{C},i_{\mathrm{S}})$ during round $i_{\mathrm{S}}$. The Pauli Z operator (light blue tile) on the \textit{pre-measure} state is only supported on data qubits and will propagate to the next $\mathrm{I}$-SE round. We can pair this light blue tile with a reset location $\mathsf{R}_{\mathrm{Z}}(\mathrm{D},i_{\mathrm{S}})$ so that together, they are jointly measured by $\mathrm{M}_{\mathrm{Z}}(\mathrm{D},i_{\mathrm{S}})$.    (\textbf{c}) Detector induced by $\{\mathsf{R}_{\mathrm{X}}(\mathrm{A},i_{\mathrm{S}}-1), \mathsf{R}_{\mathrm{X}}(\mathrm{A},i_{\mathrm{S}})\}$. (\textbf{d}) Detector induced by $\{\mathsf{R}_{\mathrm{X}}(\mathrm{C},i_{\mathrm{S}}-1),\mathsf{R}_{\mathrm{X}}(\mathrm{C},i_{\mathrm{S}}),\mathsf{R}_{\mathrm{Z}}(\mathrm{D},i_{\mathrm{S}})\}$. Detectors are marked by green squares and their measurement locations are highlighted by pentagons.   }
    \label{fig: detectors}
\end{figure}

Without loss of generality, we consider the circuit where an $\mathrm{S}$-SE round $i_{\mathrm{S}}$ is sandwiched by $\mathrm{I}$-SE rounds. As our fold-transversal operation leaves all Z stabilizers invariant, we simply use the same Z detectors (those induced solely by Z reset locations) as the memory circuit and will elaborate on how X detectors (those induced \textit{not} solely by Z reset locations) are constructed via two representative examples (Fig.~\ref{fig: detectors}(\textbf{a}-\textbf{d})). (See \cite{supplimental} for a full description of constructed detectors. )  Consider the set of reset locations $\mathcal{R}_2=\{\mathsf{R}_{\mathrm{X}}(\mathrm{A},i_{\mathrm{S}}-1),\mathsf{R}_{\mathrm{X}}(\mathrm{A},i_{\mathrm{S}})\}$, where $\mathsf{R}_{\mathrm{P}}(\mathrm{Q},i)$ is used to denote the Pauli $\mathrm{P}$-basis reset location on qubit $\mathrm{Q}$ at the beginning of  round $i$ (Fig.~\ref{fig: detectors}(\textbf{a})). During round $i_{\mathrm{S}}-1$, the X stabilizer created by $\mathsf{R}_{\mathrm{X}}(\mathrm{A},i_{\mathrm{S}}-1)$ propagates to a five-qubit flag stabilizer on the \textit{pre-measure} state. The measurement $\mathrm{M}_{\mathrm{X}}(\mathrm{A},i_{\mathrm{S}}-1)$ with result $s\in\{0,1\}$ absorbs $(-1)^{s}\mathrm{X}$ from the flag stabilizer and results in a plaquette stabilizer at the end of round $i_{\mathrm{S}}-1$. (Here $\mathrm{M}_{\mathrm{P}}(\mathrm{Q},i)$ denotes a $\mathrm{P}$-basis measurement location on qubit $\mathrm{Q}$ at the end of round $i$.) Then at the beginning of round $i_{\mathrm{S}}$, the propagated stabilizer induced by $\mathcal{R}_2$ is the flag stabilizer with sign $(-1)^{s}$, which propagates through our S gate protocol and gets jointly measured by $\mathrm{M}_{\mathrm{X}}(\mathrm{A},i_{\mathrm{S}})$ and $\mathrm{M}_{\mathrm{Z}}(\mathrm{B},i_{\mathrm{S}})$ (Fig.~\ref{fig: detectors}(\textbf{a})).  Thus $\mathcal{R}_2$ gets jointly measured by $\{\mathrm{M}_{\mathrm{X}}(\mathrm{A},i_{\mathrm{S}}-1),\mathrm{M}_{\mathrm{X}}(\mathrm{A},i_{\mathrm{S}}),\mathrm{M}_{\mathrm{Z}}(\mathrm{B},i_{\mathrm{S}})\}$, which is a detector (Fig.~\ref{fig: detectors}(\textbf{c})). 
Extra complications may arise for reset locations near the boundary, as suggested by the difference between the stabilizer patterns on the \textit{post-S} state in Fig.~\ref{fig: mid cycle states and s gate protocol}(\textbf{c}) and the \textit{half-cycle} state in Fig.~\ref{fig: mid cycle states and s gate protocol}(\textbf{a}) along the top and bottom boundary. For instance, different from the above example, reset locations $\mathsf{R}_{\mathrm{X}}(\mathrm{C},i_{\mathrm{S}}-1)$ and $\mathsf{R}_{\mathrm{X}}(\mathrm{C},i_{\mathrm{S}})$ alone do not induce a detector since the stabilizer generated by them propagates to the $i_{\mathrm{S}}+1$ round (Fig.~\ref{fig: detectors}(\textbf{b})). To obtain a detector that does not span across more than two SE rounds, we pair these two reset locations with another reset location $\mathsf{R}_{\mathrm{Z}}(\mathrm{D},i_{\mathrm{S}})$ such that $\mathcal{R}_{3}:=\{\mathsf{R}_{\mathrm{X}}(\mathrm{C},i_{\mathrm{S}}-1),\mathsf{R}_{\mathrm{X}}(\mathrm{C},i_{\mathrm{S}}),\mathsf{R}_{\mathrm{Z}}(\mathrm{D},i_{\mathrm{S}})\}$ induces a detector shown in Fig.~\ref{fig: detectors}(\textbf{d}).  Overall, we can make each X detector be contained in two adjacent SE rounds and composed of two X measurement locations on the same ancilla qubit and at most one Z measurement location.  

\textit{Decoding strategy.} Compared to the memory, circuits that contain $\mathrm{S}$-SE rounds have the following two additional types of error mechanisms that need to be addressed during decoding:
\begin{itemize}
    \item [1.] X errors prior to the S gate protocol propagate to additional Z errors. Notice that in our protocol, an X error on a data qubit may first propagate to an ancilla qubit before the fold-transversal operation and the errors on the \textit{post-S} state continue to propagate through the 3, 4 CNOT layers. 
    \item [2.] Hypergraph errors. For instance, for a detector composed of three measurement locations-two X measurements and one Z measurement, a measurement error on the latter would trigger this detector as well as two other Z detectors (Fig.~\ref{fig: detectors}(\textbf{c}-\textbf{d})). 
\end{itemize}
Our decoding strategy deals with the above complications by leveraging the fact that the Z detectors are not affected by the fold-transversal operation and that restricting to only Z detectors, both types of errors above are graph-like (or can be decomposed to graph-like ones). More concretely, we first decode solely on all Z detectors and use the decoding result to infer the correction on the X detectors and the logical observable. Then, by decoding on the corrected X detectors, we can finally obtain a collection of errors that match all syndromes and can decide whether there is a logical error. (See SM~\cite{supplimental} for a detailed description of our decoding strategy.) We notice that our decoding strategy is based on essentially the same principles as the strategies recently proposed for decoding transversal CNOT gates~\cite{wan2024iterativetransversalcnotdecoder,sahay2024errorcorrectiontransversalcnot}. Nonetheless, the setting of our decoding problem is different from these works in that we have mid-cycle X to Z error propagation on the same patch. We later develop further refinements on our decoding strategy to counter the elevation of logical error rates when $\mathrm{S}$-SE rounds are near time boundaries.

\begin{figure}[t]
    \centering
    \includegraphics[width=\columnwidth]{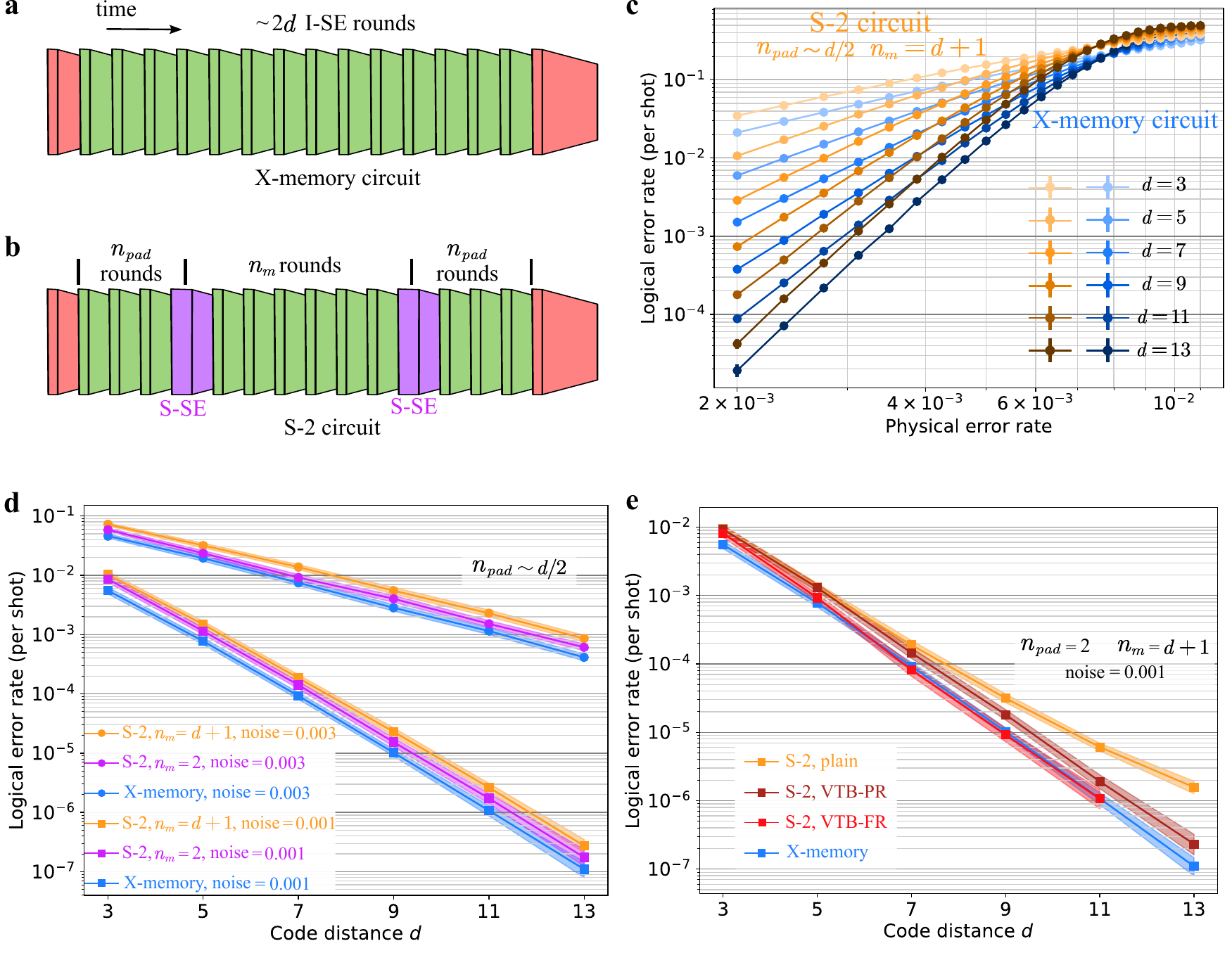}
    \caption{Decoding of the S gate protocol. (\textbf{a}) X-memory circuits used as a baseline. For code distance $d$, the X-memory circuit consists of an X-basis initialization (the left red block), $\sim 2d$ $\mathrm{I}$-SE rounds (green blocks) and a final X-basis measurement (the right red block). Here, the X-basis initialization is simply to reset all data qubits to $|+\rangle$ and the X-basis measurement is to measure all data qubits in the X basis. (\textbf{b}) S-2 circuits for benchmarking our $\mathrm{S}$ gate protocol. An S-2 circuit has two $\mathrm{S}$-SE rounds (purple blocks) inserted among $\mathrm{I}$-SE rounds and also has X-basis initialization and X-basis measurement. There are $n_{pad}$ $\mathrm{I}$-SE rounds between X-basis initialization and the first $\mathrm{S}$-SE round and another $n_{pad}$ $\mathrm{I}$-SE rounds  between the second $\mathrm{S}$-SE round and the X-basis measurement. The number of rounds between two $\mathrm{S}$-SE rounds is denoted by $n_{m}$ which is chosen to be either $\sim d$ or $2$ in our numerical experiments. (\textbf{c}-\textbf{d}) Logical error rates of the X-memory circuits (blue) and the $\mathrm{S}$-2 circuits ($n_{pad}\sim d/2$) with $n_m=d+1$ (orange) and $n_{m}=2$ (magenta) respectively, for $d$ ranging from $3$ to $13$ under various circuit error rates. (\textbf{e}) Refined decoding for $\mathrm{S}$-SE rounds near time boundaries. The $\mathrm{S}$-2 circuits with $n_{pad}=2$ and $n_{m}=d+1$ are decoded via the following three slightly different decoders: the plain decoder used in (\textbf{c}-\textbf{d}), the decoder assisted by the VTB-PR combination and the decoder assisted by the VTB-FR combination. The results for the X-memory circuits from (\textbf{d}) are reused here as a baseline. (Notice that for each $d$, the X-memory circuit is about two times longer than the $\mathrm{S}$-2 circuit used in (\textbf{e}).)  Each data point is obtained via maximum likelihood estimation over collected samples~\cite{gidney_sinter}. Error bars and shaded regions both indicate logical error rate  hypotheses whose likelihood is within a factor 1000 of the maximum likelihood.  }
    \label{fig: numerical results}
\end{figure}

\textit{S gate Performance under circuit noise.} We benchmark our S gate protocol against the memory circuits under circuit noise. The noise model used in this work is the standard circuit noise model (the SD6 model in~\cite{gidney_fault_tolerant_honeycomb_2021}) combined with pre-round depolarization error on data qubits~\cite{supplimental}. The memory circuits we use as the baseline are the X-memory circuits in Fig.~\ref{fig: numerical results}(\textbf{a}) with X-basis initialization and final measurement and $\sim 2d$ $\mathrm{I}$-SE rounds.  The circuits (the $\mathrm{S}$-2 circuits shown in Fig.~\ref{fig: numerical results}(\textbf{b})) used to benchmark our $\mathrm{S}$ gate protocol contain two $\mathrm{S}$-SE rounds separated by $n_{m}$ $\mathrm{I}$-SE rounds and also have X-basis initialization and measurement with $n_{pad}$ $\mathrm{I}$-SE rounds before the first and after the second S-SE round. We first examine the case where $\mathrm{S}$-SE rounds are kept away from time boundaries by setting $n_{pad}\sim d/2$ and experimented on two settings where $n_{m}$ is $\sim d$ and $2$ respectively. As shown in Fig.~\ref{fig: numerical results}
(\textbf{c}-\textbf{d}), our $\mathrm{S}$ gate protocol (for $n_{m}\sim d$) has similar performance as the X-memory circuits in terms of threshold and sub-threshold logical error rates. In particular, the logical error rate our $\mathrm{S}$ gate protocol is \textit{consistently} 2 to 3 times higher than the X memory circuit as we increase the code distance. Moreover, for $n_{m}=2$, we see that the logical error rate is even lower than the $n_{m}\sim d$ setting, likely due to the reduced number of $\mathrm{I}$-SE rounds (Fig.~\ref{fig: numerical results}
(\textbf{d})). Thus, even $O(1)$ rounds between two $\mathrm{S}$-SE rounds do not degrade the logical performance significantly. 

Now, we consider the case where $\mathrm{S}$-SE rounds are placed near time boundaries by setting $n_{pad}=2$.  The lack of Z detectors at the X-basis initialization or measurement round can make decoding and inference of $\mathrm{X}$ error propagation and hypergraph errors less reliable. In particular, the proximity of time boundaries incentivizes the decoder to incorrectly connect $\mathrm{Z}$ syndromes to the time boundaries thereby resulting in inference errors. (Notice that with our decoding strategy, we decode on Z detectors first and then on X detectors. Thus the decoding strategy favors results with the smallest weight for producing the Z syndromes alone but not the ones with the smallest weight for producing both Z and X syndromes. The latter results are the desired decoding results while the former are the ones obtained through our decoding strategy and may lead to higher logical error rates. We refer to the decoding on all Z (X) detectors as the Z (X) decoding step.)  As shown in Fig.~\ref{fig: numerical results}, with $n_{pad}=2$, our decoding strategy above becomes less effective at larger distances. To address this issue, we deploy two consecutive refinements on our decoding strategy (see~\cite{supplimental} for details). Inspired by the following work~\cite{gidney_yoked_2023} where different virtual spatial boundary conditions are applied to estimate the confidence in the decoding results, our first refinement is to set and iterate over four possible virtual time boundary (VTB) conditions for decoding and then select the decoding result with the smallest total weight as the final decoding result. To build on the first refinement, our second refinement aims to correlate the Z decoding step with the X decoding step. More concretely, following the idea in~\cite{fowler_optimal_2013}, at each shot, we use the decoding result in the Z decoding step to reweight the decoding graph for the X decoding step. We can either postpone the reweighting step after the X decoding step to obtain a more accurate estimate on the total weight by taking error correlation into account, or to implement the reweighting step prior to the X decoding step as in~\cite{fowler_optimal_2013}. We refer to the former as the postponed-reweighting (PR) procedure and the latter as the full-reweighting (FR) procedure.  We see in Fig.~\ref{fig: numerical results} that both VTB-PR and VTB-FR combinations are effective and the latter even manages to suppress the logical error rate down to the memory baseline.

\textit{Discussion and outlook.} Our results indicate all Clifford gates on rotated surface codes can be implemented transversally in a hardware efficient manner with reconfigurable neutral atom arrays.  As a natural next step, it would be interesting to see whether we can combine the decoder for transversal $\mathrm{S}$ gates in this work with the decoders for transversal CNOT gates~\cite{wan2024iterativetransversalcnotdecoder,sahay2024errorcorrectiontransversalcnot,wan_constant_time_2024} to efficiently decode more general Clifford circuits on rotated surface codes and whether such combination, along with our refinements for decoding logical gates near time boundaries, can achieve fast logical processing~\cite{zhou_algorithmic_2024}.    

Our \textit{mid-cycle} $\mathrm{S}$ gate protocol crucially relies on the time-dynamics of the state over both data and ancilla qubits during an SE round. The \textit{morphing} effect of SE circuits was originally discovered in~\cite{mcewen_relaxing_2023} and has been applied in designing more hardware-friendly SE circuits~\cite{mcewen_relaxing_2023,gidney_circuits_color_code_2023,shaw_lowering_bb_code_2024,debroy_luci_2024}.
Recently, a few works explore its implication in logical operations, such as extracting the logical Y operator of rotated surface codes~\cite{gidney_inplace_2024} and lattice surgery~\cite{shaw_lowering_bb_code_2024}. Our result further extends the potential of the \textit{morphing} effect by using it to construct a transversal $\mathrm{S}$ gate on rotated surface codes. It is an interesting future direction to consider enlarging the transversal gate sets of certain qLDPC codes, such as the Hypergraph Product codes~\cite{tillich_quantum_2014,quintavalle_partitioning_2023,xu_fast_2024} and the Bivariate-Bicycle codes~\cite{bravyi_high_threshold_2024,eberhardt_logical_2024,shaw_lowering_bb_code_2024}, by \textit{morphing} the data and ancilla qubits to other qLDPC codes with a different set of transversal gates during SE. More generally, we expect to see more efficient FTQC strategies constructed from circuit-level perspectives~\cite{mcewen_relaxing_2023,gottesman_opportunities_2022,delfosse_spacetime_2023,beverland_fault_2024}.

\textit{Code availability.} We used Stim~\cite{gidney2021stim} for noisy circuit sampling and construction of detector error models, Pymatching~\cite{higgott_sparse_2023} for decoding, and Sinter~\cite{gidney_sinter} for statistical analysis of the logical error rates. The codes and circuits for our numerical results are publicly available at~\cite{zihan_dynamical_phase_gate}.

\bibliography{References}

\end{document}


\title{Supplementary Information: Transversal Logical Clifford gates on rotated surface codes with reconfigurable neutral atom arrays} 

\author{Zihan Chen}


\affiliation{
Hefei National Research Center for Physical Sciences at the Microscale and School of Physical Sciences,
University of Science and Technology of China, Hefei 230026, China}
\affiliation{
Shanghai Research Center for Quantum Science and CAS Center for Excellence in Quantum Information and Quantum Physics,
University of Science and Technology of China, Shanghai 201315, China}
\affiliation{
Hefei National Laboratory, University of Science and Technology of China, Hefei 230088, China}

\author{Mingcheng Chen}

\affiliation{
Hefei National Research Center for Physical Sciences at the Microscale and School of Physical Sciences,
University of Science and Technology of China, Hefei 230026, China}
\affiliation{
Shanghai Research Center for Quantum Science and CAS Center for Excellence in Quantum Information and Quantum Physics,
University of Science and Technology of China, Shanghai 201315, China}
\affiliation{
Hefei National Laboratory, University of Science and Technology of China, Hefei 230088, China}

\author{Chaoyang Lu}

\affiliation{
Hefei National Research Center for Physical Sciences at the Microscale and School of Physical Sciences,
University of Science and Technology of China, Hefei 230026, China}
\affiliation{
Shanghai Research Center for Quantum Science and CAS Center for Excellence in Quantum Information and Quantum Physics,
University of Science and Technology of China, Shanghai 201315, China}
\affiliation{
Hefei National Laboratory, University of Science and Technology of China, Hefei 230088, China}

\author{Jianwei Pan}

\affiliation{
Hefei National Research Center for Physical Sciences at the Microscale and School of Physical Sciences,
University of Science and Technology of China, Hefei 230026, China}
\affiliation{
Shanghai Research Center for Quantum Science and CAS Center for Excellence in Quantum Information and Quantum Physics,
University of Science and Technology of China, Shanghai 201315, China}
\affiliation{
Hefei National Laboratory, University of Science and Technology of China, Hefei 230088, China}

\date{\today}

\maketitle

\section{Detailed description of the mid-cycle S gate protocol}

\subsection{Notations and framework for detector construction}

\begin{figure*}[t]
    \centering
    \includegraphics[width=\textwidth]{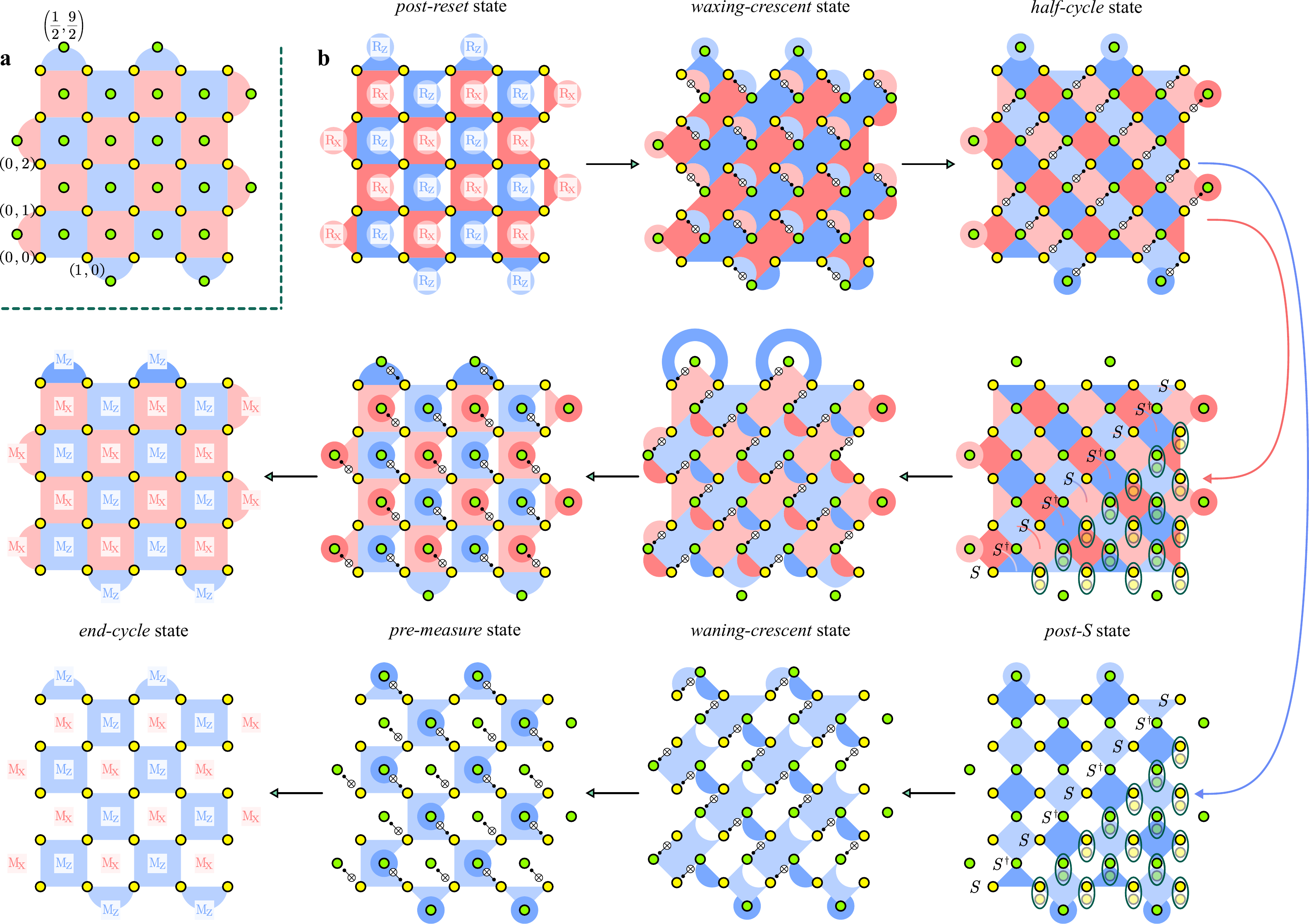}
    \caption{Stabilizer trajectories and circuit-level operations during an $\mathrm{S}$-SE round. (\textbf{a}) Coordinate system for data and ancilla qubits. (\textbf{b}) Stabilizer trajectories generated by reset locations $\mathsf{R}_{\mathrm{P}}(x,y,i_{\mathrm{S}}-1)$ and $\mathsf{R}_{\mathrm{P}}(x,y,i_{\mathrm{S}})$ during an $\mathrm{S}$-SE round $i_{\mathrm{S}}$. The red (blue) arrow connecting the \textit{half-cycle} state to the \textit{post-S} state indicates the propagation of stabilizers induced by $\mathrm{X}$-reset ($\mathrm{Z}$-reset) locations. Illustrated operations (including gates, resets and measurements) on a \textit{mid-cycle} state represents the physical operations immediately before. }
    \label{suppfig: SSE full round}
\end{figure*}

We introduce notations convenient for our discussions later. As in Fig.~\ref{suppfig: SSE full round}(\textbf{a}), we set a coordinate system on the rotated surface code to describe data qubits (with integer coordinates) and ancilla qubits (with half-integer coordinates). A single-qubit Pauli operator $\mathrm{P}$ on qubit $(x,y)$ is denoted as $\mathrm{P}_{(x,y)}$; the restriction of a multi-qubit Pauli operator $\mathcal{P}$ on qubit $(x,y)$ is denoted as $\mathcal{P}|_{(x,y)}$. During an SE round, we refer to the state over all data qubits and ancilla qubits between two adjacent layers of physical operations as a \textit{mid-cycle} state and label it as in Fig.~\ref{suppfig: SSE full round}(\textbf{b}), following much of the convention in~\cite{mcewen_relaxing_2023}. Given a \textit{mid-cycle} state lable \textit{mc} as in Fig.~\ref{suppfig: SSE full round}(\textbf{b}) and an SE round index $i$, we can use $(mc,i)$ as a timestamp and use $(mc,i)\preceq (\overline{mc},j)$ to indicate timestamp $(\overline{mc},j)$ is not earlier than $(mc,i)$ and similarly $\prec$ is used to indicate strict temporal precedence. For a stabilizer $\mathrm{S}$ on the \textit{mc} state at round $i$, we define the corresponding stabilizer location as $\mathsf{S}:=(\mathrm{S},(mc,i))$. To simplify notation, we assume that the measurement locations in each round $i$ are all single-qubit Pauli-basis measurements that only occur between $(\textit{pre-measure},i)$ and $(\textit{end-cycle},i)$; and that the reset locations only occur at the beginning of each round. Moreover, we assume that measured qubits no longer participate in the circuit unless they are reset.

To keep track of the stabilizer propagation, for each stabilizer location $\mathsf{S}$, we define its \textit{trajectory} $\mathcal{T}(\mathsf{S})$ to register the propagation of $\mathsf{S}$ such that $\mathcal{T}(\mathsf{S},(\overline{\textit{mc}},j))$ is the (forward- or backward-) propagated stabilizer at time-step $(\overline{\textit{mc}},j)$. We identify a $\mathrm{P}$-basis reset location $\mathsf{R}_{\mathrm{P}}(x,y,i)$ on qubit $(x,y)$ at the beginning of round $i$ as a special stabilizer location whose backward propagation is defined to be the identity operator. In other words, the stabilizer trajectory $\mathcal{T}(\mathsf{R}_{\mathrm{P}}(x,y,i))$ satisfies $\mathcal{T}(\mathsf{R}_{\mathrm{P}}(x,y,i),(\overline{mc},j))=1$ for $(\overline{mc},j)\prec (\textit{post-reset},i)$ and $\mathcal{T}(\mathsf{R}_{\mathrm{P}}(x,y,i),(\textit{post-reset},i))=\mathrm{P}_{(x,y)}$. We describe in the following how measurements affect stabilizer trajectories (tracking stabilizers through unitary channels is more straight-forward). Consider a $\mathrm{P}$-basis measurement location $\mathrm{M}_{\mathrm{P}}(x,y,i)$ on qubit $(x,y)$ in round $i$,  and a stabilizer trajectory $\mathcal{T}(\mathsf{S})$, we say $\mathrm{M}_{\mathrm{P}}(x,y,i)$ \textit{absorbs} from $\mathcal{T}(\mathsf{S})$ if $\mathcal{T}(\mathsf{S},(\textit{pre-measure},i))|_{(x,y)}=\mathrm{P}$ and \textit{annihilates} $\mathcal{T}(\mathsf{S})$ if $\mathcal{T}(\mathsf{S},(\textit{pre-measure},i))$ anticommutes with $\mathrm{P}_{(x,y)}$. If $\mathcal{T}(\mathsf{S})$ gets \textit{absorbed} by $\mathrm{M_P}(x,y,i)$ with measurement outcome $s\in\{0,1\}$, $\mathcal{T}(\mathsf{S},(\textit{pre-measure},i))$ would propagate to 
\[\mathcal{T}(\mathsf{S},(\textit{end-cycle,i}))=(-1)^{s}\mathrm{P}_{(x,y)}\mathcal{T}(\mathsf{S},(\textit{pre-measure}),i).\]
On the other hand, if $\mathcal{T}(\mathsf{S})$ gets \textit{annihilated} by $\mathrm{M_P}(x,y,i)$, $\mathcal{T}(\mathsf{S},(\textit{pre-measure,i}))$ will not propagate to the next time step and naturally $\mathcal{T}(\mathsf{S},(\overline{mc},j))$ is set to $0$ for $(\overline{mc},j)\succeq (\textit{end-cycle},i)$.  Given the above structures of stabilizer trajectories, we can naturally define the product (denoted by $\cdot$) of two stabilizer trajectories $\mathcal{T}(\mathsf{S}_{1})$ and $\mathcal{T}(\mathsf{S}_{2})$
as another stabilizer trajectory $\mathcal{T}(\{\mathsf{S}_1,\mathsf{S}_2\}):=\mathcal{T}(\mathsf{S}_{1})\cdot\mathcal{T}(\mathsf{S}_{2})$ with $\mathcal{T}(\{\mathsf{S}_1,\mathsf{S}_2\},(\overline{mc},j))=\mathcal{T}(\mathsf{S}_1,(\overline{mc},j))\cdot\mathcal{T}(\mathsf{S}_2,(\overline{mc},j))$ for any timestamp $(\overline{mc},j)$. Then, given a set of stabilizer locations $\{\mathsf{S}_{i}\}$, we can define the stabilizer trajectory generated by all stabilizer locations in this set as $\mathcal{T}(\{\mathsf{S}_{i}\}):=\prod_{i}\mathcal{T}(\mathsf{S}_i)$.

Using the above language, we formulate how detectors are constructed in the following. Our discussion is  based on the detecting region formalism~\cite{mcewen_relaxing_2023} and will be more explicit and formal on the construction of detectors and their corresponding detecting regions. We say a stabilizer trajectory $\mathcal{T}(\mathsf{S})$ is \textit{fully absorbed} by a set of (single-qubit) measurements $\mathcal{M}=\{\mathrm{M}_{\mathrm{P}}(x,y,i)\}$ if  $\mathcal{T}(\mathsf{S},(\textit{end-cycle},i_{max}))$ is reduced to $1$ (where $i_{max}$ is the maximal round in $\mathcal{M}$) and $\mathcal{M}$ is the set of all measurement locations that \textit{absorb} from $\mathcal{T}(\mathsf{S})$. Then, it is straightforward to show that given a set of reset locations $\mathcal{R}=\{\mathsf{R}_{\mathrm{P}}(x,y,i)\}$, if the stabilizer trajectory generated by $\mathcal{R}$ is \textit{fully absorbed} by a set of measurements, the parity sum of these measurement results is a detector. Moreover, such generated stabilizer trajectory is exactly a detecting region. Conversely, given a detector, by back-propagating the jointly-measured Pauli operators, we can obtain its corresponding detecting region as well as the reset locations that generate the detection region.~\cite{mcewen_relaxing_2023} For constructing detectors, we use the forward-propagating perspective described above. The framework of detecting regions will be used later to analyse \textit{in situ} how error locations trigger detectors without having to track how these errors propagate.

\subsection{Stabilizer trajectories during an S-SE round and detectors}

\newcommand{\commabreak}{,\allowbreak}
We explicitly illustrate the operations and stabilizer trajectories during an S-SE round in Fig.~\ref{suppfig: SSE full round}(\textbf{b}). For a circuit that sandwiches a single $\mathrm{S}$-SE round $i_{\mathrm{S}}$ with SE rounds, all constructed detectors along with their assigned coordinate labels are listed as follows. 
\begin{itemize}
    \item [1.] Z-detectors. For each $\mathsf{R}_\mathrm{Z}(x,y,i)$, $\{\mathsf{R}_\mathrm{Z}(x\commabreak y\commabreak i)\commabreak \mathsf{R}_\mathrm{Z}\allowbreak(x\commabreak y\commabreak i+1)\}$ induces the detector $\mathrm{D}_\mathrm{Z}(x,y,i+\frac{1}{2}):=\{\mathrm{M}_\mathrm{Z}(x,y,i),$ 
 $\mathrm{M}_\mathrm{Z}(x,y,i+1)\}$. 
    \item  [2.] X-detectors. For each $\mathsf{R}_\mathrm{X}(x,y,i)$,
        \begin{itemize}
            \item [i.] if $i\leq i_{\mathrm{S}}-2$ or $i\geq i_{\mathrm{S}}+1$, $\{\mathsf{R}_\mathrm{X}(x,\allowbreak y,\allowbreak i),$ $\mathsf{R}_\mathrm{X}(x,y,i+1)\}$ induces the detector $\mathrm{D}_{\mathrm{X}}(x,y,i+\frac{1}{2}):=\{\mathrm{M}_\mathrm{X}(x,\allowbreak y,\allowbreak i),$ $\mathrm{M}_\mathrm{X}(x,y,i+1)\}$;
            \item [ii.] if $i=i_{\mathrm{S}}-1$ and $x=d+\frac{1}{2}$, $\{\mathsf{R}_\mathrm{X}(x,\allowbreak y,\allowbreak i),$ $\mathsf{R}_\mathrm{X}(x,y,i+1)\}$ induces the detector $\mathrm{D}_{\mathrm{X}}(x,y,i+\frac{1}{2}):=\{\mathrm{M}_\mathrm{X}(x,\allowbreak y,\allowbreak i),$ $\mathrm{M}_\mathrm{X}(x,y,i+1)\}$;
            \item [iii.] if $i=i_{\mathrm{S}}-1$ and $x< d-\frac{1}{2}$, $\{\mathsf{R}_\mathrm{X}(x\commabreak y\commabreak i)\commabreak \mathsf{R}_\mathrm{X}(x\commabreak y\commabreak i+1)\}$ induces the detector $\mathrm{D}_{\mathrm{X}}(x,y,i+\frac{1}{2}):=\{\mathrm{M}_\mathrm{X}(x\commabreak y\commabreak i)\commabreak \mathrm{M}_\mathrm{X}(x\commabreak y\commabreak i+1)\commabreak \mathrm{M}_\mathrm{Z}(y\commabreak x+1\commabreak i+1)\}$;
            \item [iv.] if $i=i_{\mathrm{S}}-1$ and $x=d-\frac{1}{2}$, $\{\mathsf{R}_\mathrm{X}(x\commabreak y\commabreak i)\commabreak \mathsf{R}_\mathrm{X}(x\commabreak y\commabreak i+1)\commabreak \mathsf{R}_\mathrm{Z}(y\commabreak x+1\commabreak i+1)\}$ induces the detector $\mathrm{D}_{\mathrm{X}}(x,y,i+\frac{1}{2}):=\{\mathrm{M}_\mathrm{X}(x\commabreak y\commabreak i)\commabreak \mathrm{M}_\mathrm{X}(x\commabreak y\commabreak i+1)\commabreak \mathrm{M}_\mathrm{Z}(y\commabreak x+1\commabreak i+1)\}$;
            \item [v.] if $i=i_{\mathrm{S}}$ and $x=-\frac{1}{2}$, $\{\mathsf{R}_\mathrm{X}(x\commabreak y\commabreak i)\commabreak \mathsf{R}_\mathrm{X}(x\commabreak y\commabreak i+1)\}$ induces the detector $\{\mathrm{M}_\mathrm{X}(x\commabreak y\commabreak i)\commabreak \mathrm{M}_\mathrm{X}(x\commabreak y\commabreak i+1)\}$;
            \item [vi.] if $i=i_{\mathrm{S}}$ and $x=\frac{1}{2}$, $\{\mathsf{R}_\mathrm{X}(x\commabreak y\commabreak i)\commabreak \mathsf{R}_\mathrm{X}(x\commabreak y\commabreak i+1)\commabreak \mathsf{R}_\mathrm{Z}(y\commabreak x-1,i+1)\}$ induces the detector $\mathrm{D}_{\mathrm{X}}(x,y,i+\frac{1}{2}):=\{\mathrm{M}_\mathrm{X}(x\commabreak y\commabreak i)\commabreak \mathrm{M}_\mathrm{X}(x\commabreak y\commabreak i+1)\commabreak \mathrm{M}_\mathrm{Z}(y\commabreak x-1\commabreak i+1)\}$;
            \item [vii.] if $i=i_{\mathrm{S}}$ and $x>\frac{1}{2}$, $\{\mathsf{R}_\mathrm{X}(x\commabreak y\commabreak i)\commabreak \mathsf{R}_\mathrm{X}(x\commabreak y\commabreak i+1)\commabreak \mathsf{R}_\mathrm{Z}(y\commabreak x-1\commabreak i+1)\}$ induces the detector $\mathrm{D}_{\mathrm{X}}(x,y,i+\frac{1}{2}):=\{\mathrm{M}_\mathrm{X}(x\commabreak y\commabreak i)\commabreak \mathrm{M}_\mathrm{X}(x\commabreak y\commabreak i+1)\}$, where we implicitly used the fact that $\{\mathrm{M}_\mathrm{Z}(y\commabreak x-1\commabreak i)\commabreak \mathrm{M}_\mathrm{Z}(y\commabreak x-1\commabreak i+1)\}$ is already a detector. 
         \end{itemize}
\end{itemize}

\begin{figure*}[t]
    \centering
    \includegraphics[width=\textwidth]{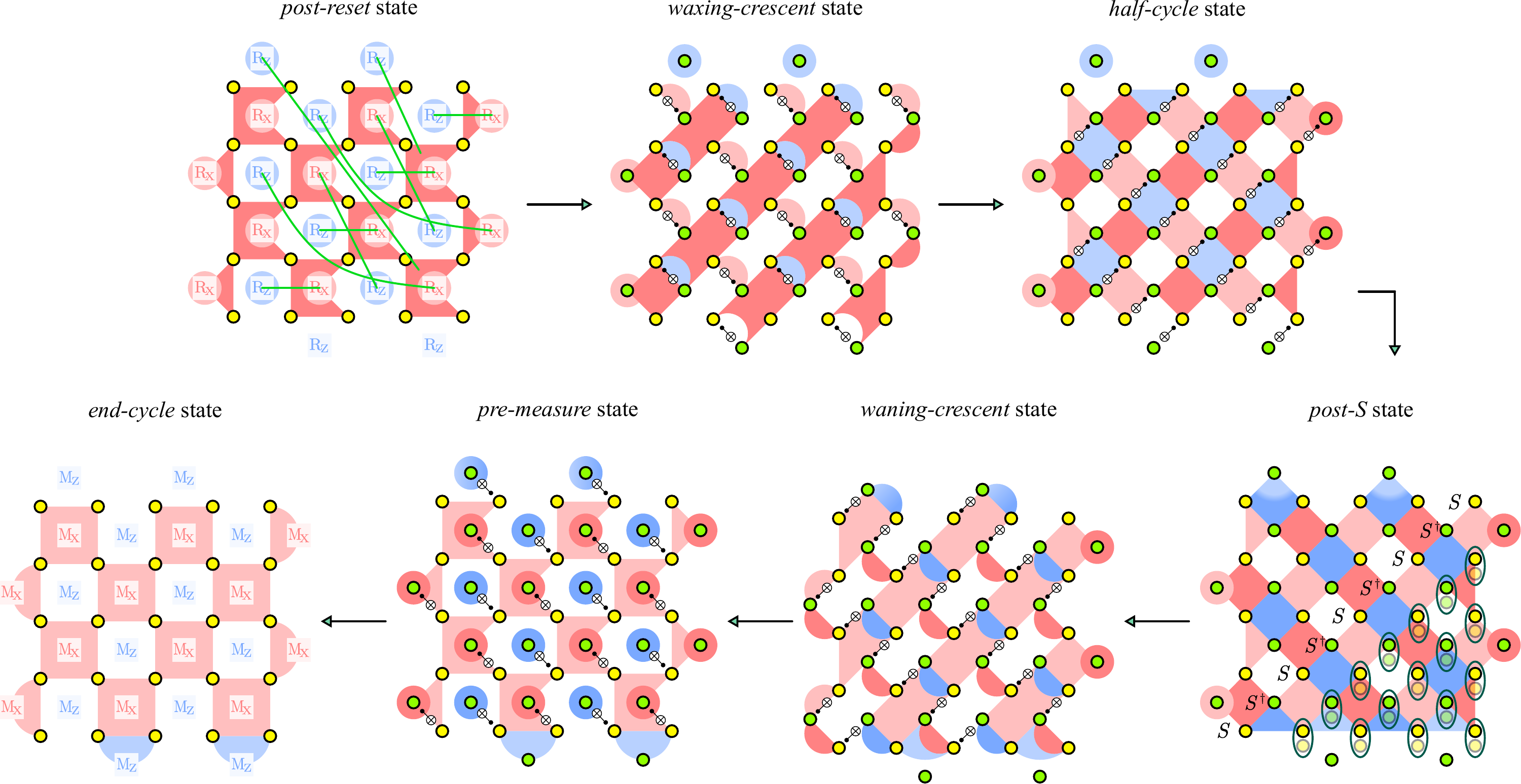}
    \caption{Detecting regions associated with the X-detectors during an $\mathrm{S}$-SE round. Reset locations corresponding to the same  X-detector are linked by green lines on the \textit{post-reset} state. Red flag tiles along the right boundary are paired with light blue tiles on the top boundary. Such pairings result in blended blue tiles on the top boundary for the \textit{post-S}, \textit{waning-crescent} and \textit{pre-measure} state.}
    \label{suppfig: x detecting regions sse}
\end{figure*}

\section{Decoding} 

\subsection{Error model}

We use the SD6 circuit noise model~\cite{gidney_fault_tolerant_honeycomb_2021} combined with pre-round depolarization noise in our simulation. More specifically, our circuits have the following types of error locations:
\begin{itemize}
    \item [i.] each single- (two-) qubit gate (including idling) is followed by a single- (two-) qubit depolarization channel with noise strength $p$,
    \item [ii.] error rate for each reset and measurement location is $p$ and
    \item  [iii.] each data qubit at the begining of each SE or $\mathrm{S}$-SE round is subject to a single-qubit depoliarziation noise with noise strength $p$. 
\end{itemize}

Our circuits start with an $\mathrm{X}$-basis initialization which is implemented by initializing all data qubits to $|+\rangle$, and end with an X-basis measurement which is implemented by measuring all data qubits in the $\mathrm{X}$ basis. We denote our logical outcome by $L$, which should have a fixed value for both a noiseless X-memory circuit and a noiseless $\mathrm{S}$-2 circuit. 

\subsection{Decoding strategy}\label{suppsubsec: decoding strategy}

We start by describing the construction and structure of our decoding hypergraph $\mathcal{G}(\mathcal{D},\mathcal{E})$ (or $\mathcal{G}$ for short) whose vertex set $\mathcal{D}$ is the set of all detectors and hyperedge set $\mathcal{E}$ is a set of error locations to be specified later. (Here $\mathcal{E}$ is a subset of the collection of all error locations, $\mathcal{E}_{tot}$.) Denote the set of detectors triggered by an error location $e$ as $\partial e$ and its restriction to the X (Z) detectors as $\partial_{\mathrm{X}}e$ ($\partial_{\mathrm{Z}}e$).  In the decoding hypergraph $\mathcal{G}(\mathcal{D},\mathcal{E})$, for each error location $e\in\mathcal{E}$, the collection of its endpoints is simply $\partial e$. To make sure our decoding problem is amenable to matching-based decoders, $\mathcal{E}$ is chosen to be the maximal subset of the set of all error locations such that each $e\in\mathcal{E}$ is a (single- or two-qubit) X or Z error location that satisfies both
\begin{equation} 
\label{suppeq: error egde criterion}
|\partial_{\mathrm{X}}e|\leq 2 \quad \text{and}\quad  |\partial_{\mathrm{Z}}e|\leq 2,
\end{equation}  
and is unique in the sense that there does not exist another edge $e'\in\mathcal{E}$ with $\partial e'=\partial e$ and the same logical effect as $e$. (For convenience, we assume in the following without loss of generality that each error location is unique.) 

To efficiently readout the triggered detectors by an error location $e$, the detecting region formalism is used to circumvent the need to forward-propagate the error location~\cite{mcewen_relaxing_2023}. More specifically, given a detector $\mathrm{D}$ induced by a set of reset locations $\mathcal{R}$, an error location $e:=(\mathrm{P},(\textit{mc},i))$ triggers $\mathrm{D}$ iff $\mathrm{P}$ anti-commutes with $\mathcal{T}(\mathcal{R},(\textit{mc},i))$.   As any error location can be decomposed into single-qubit X and Z error locations, we first discuss how a single-qubit X (Z) error location may trigger detectors. For Z detectors, we see that each qubit at each time step intersects with at most two Z detecting regions, thus each single-qubit X error at most flip two Z detectors. For X detectors, since X detecting regions in $\mathrm{I}$-SE rounds are always represented by red tiles (X operators), a single-qubit X error location during an $\mathrm{I}$-SE round does not flip any X detector. On the other hand, during an S-SE round, as shown in Fig.~\ref{suppfig: x detecting regions sse}, an X detecting region may contain a blue tile, and each qubit is on at most two such blue tiles (except for certain qubits near the boundary, most qubits are on only one blue tile). Thus, a single-qubit X error location during an S-SE round may flip at most two X detectors. (The majority of single-qubit X error locations during an $\mathrm{S}$-SE round flip zero or one X detector.) To summarize, a single-qubit X error location flip at most two Z detectors and at most two X detectors. It is also straight forward to see that a single-qubit Z error location flips at most two X detectors and does not flip any Z detector. As a result, all single-qubit X (Z) errors are contained in $\mathcal{E}$, which means our decoding hypergraph $\mathcal{G}$ contains enough hyperedges to account for syndromes triggered by any collection of error locations.    
As for two-qubit X (Z) errors, X (Z) error locations associated with CNOT gates are already encompassed by single-qubit X (Z) error locations since each two-qubit X (Z) error after a CNOT gate is equivalent to a single-qubit X (Z) error immediately prior to the CNOT gate. In contrast, most two-qubit X (Z) errors associated with the CZ gates of the fold-transversal operations do not satisfy the condition in Eq.~\ref{suppeq: error egde criterion} and are not chosen as (hyper-) edges in our decoding hypergraph. 

For ordinary memory circuits, the decoding hypergraph $\mathcal{G}$ is naturally a disjoint union of its Z-subhypergraph $\mathcal{G}(\mathcal{D}_{\mathrm{Z}},\mathcal{E}_{\mathrm{Z}})$ (or $\mathcal{G}_{\mathrm{Z}}$ for short) and its X-subhypergraph $\mathcal{G}(\mathcal{D}_{\mathrm{X}},\mathcal{E}_{\mathrm{X}})$ (or $\mathcal{G}_{\mathrm{X}}$ for short) where $\mathcal{D}_{\mathrm{Z}}$ ($\mathcal{D}_{\mathrm{X}}$) is the collection of $\mathrm{Z}$ ($\mathrm{X}$) detectors and $\mathcal{E}_{\mathrm{Z}}\subset \mathcal{E}$ ($\mathcal{E}_{\mathrm{X}}\subset\mathcal{E}$) is the subset of all hyperedges that have endpoints in $\mathcal{D}_{\mathrm{Z}}$ ($\mathcal{D}_{\mathrm{X}}$). In such cases, the decoding problem can be split into (separate) decoding on $\mathcal{G}_{\mathrm{Z}}$ and on $\mathcal{G}_{\mathrm{X}}$.  However, for more interesting circuits with $\mathrm{S}$-SE rounds, there are hyperedges in $\mathcal{E}$ connecting vertices in $\mathcal{D}_{\mathrm{Z}}$ to vertices in $\mathcal{D}_{\mathrm{X}}$. We partition $\mathcal{E}$ into three disjoint subsets $\tilde{\mathcal{E}}_{\mathrm{Z}}$, $\tilde{\mathcal{E}}_{\mathrm{X}}$ and $\tilde{\mathcal{E}}_{\mathrm{mix}}$ where $\tilde{\mathcal{E}}_{\mathrm{Z}}$  ($\tilde{\mathcal{E}}_{\mathrm{X}}$) is the set of edges with  endpoints contained in $\mathcal{D}_{\mathrm{Z}}$ ($\mathcal{D}_{\mathrm{X}}$) and $\tilde{\mathcal{E}}_{\mathrm{mix}}$ is the set of hyperedges straddling $\mathcal{D}_{\mathrm{Z}}$ and $\mathcal{D}_{\mathrm{X}}$. Then, by definition, $\mathcal{E}_{\mathrm{Z}}=\tilde{\mathcal{E}}_{\mathrm{Z}}\sqcup \tilde{\mathcal{E}}_{\textit{mix}}$ and $\mathcal{E}_{\mathrm{X}}=\tilde{\mathcal{E}}_{\mathrm{X}}\sqcup \tilde{\mathcal{E}}_{\textit{mix}}$.  For each hyperedge $e\in\tilde{\mathcal{E}}_{\mathrm{mix}}$, either $|\partial_{\mathrm{Z}}e|=2$ or $\partial_{\mathrm{Z}}e$ is a boundary detector, whereas $\partial_{\mathrm{X}}e$ is typically a single vertex in the bulk. Thus, such hyperedges along with errors in $\tilde{\mathcal{E}}_{\mathrm{Z}}$ can both be identified via their footprint on $\mathcal{G}_{\mathrm{Z}}$. (There are ambiguities where two distinct hyperedges have the same footprint on $\mathcal{G}_{\mathrm{Z}}$, as also noticed for decoding transversal CNOT gates in~\cite{sahay2024errorcorrectiontransversalcnot}. However, according to our numerical results in the main text, such ambiguities do not pose a serious problem.) As for errors in $\tilde{\mathcal{E}}_{\mathrm{X}}$, we define a new decoding hypergraph $\tilde{\mathcal{G}}(\mathcal{D}_{\mathrm{X}},\tilde{\mathcal{E}}_{\mathrm{mix}})$ or $\tilde{\mathcal{G}}_{\mathrm{X}}$ for short.  Based on the above observations, given a set $\mathcal{D}_{s}$ of triggered detectors, our decoding works in the following way.
\begin{itemize}
    \item [1.] Z-decoding step. Given Z-syndromes $\mathcal{D}_{s}\cap \mathcal{D}_{\mathrm{Z}}$, decode on $\mathcal{G}_{\mathrm{Z}}$ to obtain an error set $\mathrm{E}_{\mathrm{Z}}\subset \mathcal{E}_{\mathrm{Z}}$. 
    \item [2.] Correction step. Correct the X-syndromes to $(\mathcal{D}_{s}\cap\mathcal{D}_{\mathrm{X}})\oplus \sum_{e\in\mathrm{E}_\mathrm{Z}} \partial_{\mathrm{X}}e$.
    \item [3.] X-decoding step. Decode on $\tilde{\mathcal{G}}_{\mathrm{X}}$ to obtain another error set $\mathrm{E}_{\mathrm{X}}\subset \tilde{\mathcal{E}}_{\mathrm{X}}$. (It is straight forward to verify that $\mathrm{E}_\mathrm{Z}\sqcup \mathrm{E}_\mathrm{X}$ provides the correct syndromes $\mathcal{D}_{s}$.) 
    \item  [4.] Merge step. The correction to the logical observable is $\oplus_{e\in\mathrm{E}_\mathrm{Z}\sqcup\mathrm{E}_\mathrm{X}}\allowbreak L(e)$.  Here $L(e)$ is the logical effect of $e$ on the logical outcome. 
\end{itemize}
We note that our decoding strategy above is based on essentially the same principles as the recent works on decoding CNOT gates~\cite{wan2024iterativetransversalcnotdecoder,sahay2024errorcorrectiontransversalcnot}. As explained in the main text, their decoding problem is different from ours.

\subsection{Refinements of the decoding strategy I: varying over virtual time boundary conditions}

The refinement here is inspired by the following work~\cite{gidney_yoked_2023} where artificial spatial boundary conditions of a decoding problem are set and varied to estimate the confidence in the decoding result (or to estimate the complementary gap according to the original terminology in~\cite{gidney_yoked_2023}).  In our case, we define four possible time boundary conditions for the $\mathcal{G}_{\mathrm{Z}}$ decoding graph and iterate over them to obtain a decoding result with the lowest total weight. We describe how time boundary conditions are set in the following. 

We index the X-basis initialization round by $i_{0}$ and the X-basis measurement round by $i_{f}$. The two spatial slices of Z detectors with the smallest time coordinate $i_0+\frac{1}{2}$ and with the largest time coordinate $i_{f}-\frac{1}{2}$ are denoted by  $\mathcal{D}_{\mathrm{Z}}(i_0+\frac{1}{2})$ and $\mathcal{D}_{\mathrm{Z}}(i_{f}-\frac{1}{2})$ respectively. We add two vertices $v_0$ and $v_f$ (acting as virtual detectors) to the decoding graph $\mathcal{G}_{\mathrm{Z}}$. For each edge $e\in\mathcal{E}_{\mathrm{Z}}$, if $|\partial_{\mathrm{Z}} e|=1$ and $\partial e\subset \mathcal{D}_{\mathrm{Z}}(i_0+\frac{1}{2})$, we connect $e$ to $v_0$. Similarly, if $|\partial_{\mathrm{Z}} e|=1$ and $\partial_{\mathrm{Z}}e\subset \mathcal{D}_{\mathrm{Z}}(i_{f}-\frac{1}{2})$, we connect $e$ to $v_{f}$. A time boundary condition on the modified $\mathcal{G}_{\mathrm{Z}}$ is set by fixing whether $v_{0}$ and $v_{f}$ is triggered. Given a set of syndromes $\mathcal{D}_{s}$, our refined decoding procedure works in the following  way:
\begin{itemize}
    \item [1.] For each one of the four possible time boundary conditions ($(v_0,v_f)\in\{0,1\}^{2}$), perform the decoding procedure in Sec.~\ref{suppsubsec: decoding strategy} and register the total weight of the decoding result (which contains both errors in $\mathcal{E}_{\mathrm{Z}}$ and in $\tilde{\mathcal{E}}_{\mathrm{X}}$) and the correction to the logical observable. 
    \item [2.] Select the decoding result (and the associated logical observable correction) with the smallest total weight as the final decoding result. 
\end{itemize}
Notice that in the above procedure, a straight forward way to calculate the total weight of the decoding result is by assuming errors in $\mathcal{E}_{\mathrm{Z}}$ and $\tilde{\mathcal{E}}_{X}$ are independent and simply summing over the weights of the decoding results. A more accurate estimation of total weights requires taking into account the error locations that are combinations of errors in $\mathcal{E}_{\mathrm{Z}}$ and $\tilde{\mathcal{E}}_{\mathrm{X}}$, thereby inducing correlation between error locations in $\mathcal{E}_{\mathrm{Z}}$ and in $\tilde{\mathcal{E}}_{\mathrm{X}}$. In the following, we describe another refinement that accounts for such correlation between $\mathcal{E}_{\mathrm{Z}}$ and $\tilde{\mathcal{E}}_{\mathrm{X}}$. 

\subsection{Refinements of the decoding strategy II: correlating the Z-decoding step with the X-decoding step}
Inspired by~\cite{fowler_optimal_2013}, we can use the error results from the Z-decoding step to infer how more general errors in $\mathcal{E}_{tot}$ may have happened. Based on the inference result, $\tilde{\mathcal{G}}_{\mathrm{X}}$ is be reweighted. We can either use the reweighted $\tilde{\mathcal{G}}_{\mathrm{X}}$ only after the X decoding step to recalculate the total weight of the decoding result or to carry out the X decoding step directly under the reweighted $\tilde{\mathcal{G}}_{\mathrm{X}}$. The former way is called the post-reweight (PR) procedure and the latter is called the full-reweight (FR) procedure.   We describe the inference and reweighting procedures above in detail in the following. 

Each $e\in\mathcal{E}_{tot}$ is decomposed into errors in $\mathcal{E}_{\mathrm{Z}}$ and $\tilde{\mathcal{E}}_{\mathrm{X}}$. We use $e_0\in e$ to indicate $e_0$ is contained in the decomposition of $e$ and define $N(e_0):=\{e\in\mathcal{E}_{tot}:e_0\in e\} $ as the set of errors whose decompositions contain $e_0$. Then, given a decoded error $e_{z}\in\mathcal{E}_{\mathrm{Z}}$, heuristically each error in $N(e_{z})$ may happen with probability $1/|N(e_{z})|$. (A more careful inference can also be implemented accounting for different prior probabilities for errors in $N(e_{z})$.) For a set of decoded errors $\mathcal{E}_{d}:=\{e_{i}\}\subset\mathcal{E}_{\mathrm{Z}}$, the inference procedure is as follows:
\begin{itemize}
    \item [1.] For each $e\in\mathcal{E}_{tot}$, its error probability $P_{o}(e)$ is reset to 0 iff $e\in\cup_{e_{i}\in\mathcal{E}_{d}}N(e_{i})$. 
    \item  [2.] For each $e_{i}\in\mathcal{E}_{d}$ and each $e\in N(e_{i})$, update $P_{o}(e)$ to $P_{o}(e)+\frac{1}{|N(e_{i})|}$. 
\end{itemize}
After the inference step, the error probabilities for errors in $\mathcal{E}_{tot}$ are updated. Based on this new information, the reweighting procedure in $\tilde{\mathcal{G}}_{\mathrm{X}}$ is implemented by setting the error probability of each $e_{x}\in\tilde{\mathcal{E}}_{\mathrm{X}}$ to $P(e_x)=\sum_{e\in N(e_{x})}P_{o}(e)$. Notice that we can also use the belief-propagation (BP) algorithm, following~\cite{higgott_improved_2023}, to update error probabilities for error locations in $\mathcal{E}_{tot}$ based on syndromes for each shot. We leave such implementation for future work.

\bibliography{References_SI}